\documentclass[aps,twocolumn,prd,showpacs,nofootinbib]{revtex4}
\usepackage{amsmath}
\usepackage{graphicx}
\usepackage{dcolumn}
\usepackage{bm}
\usepackage{amssymb}
\usepackage{latexsym}

\def\be{\begin{equation}}
\def\ee{\end{equation}}
\def\ba{\begin{eqnarray}}
\def\ea{\end{eqnarray}}

\begin{document}

\title{Suppressing CMB Quadrupole with a Bounce from
Contracting Phase to
Inflation}

\author{Yun-Song Piao${}^{a,b}$}
\email{yspiao@itp.ac.cn}
\author{Bo Feng${}^a$}
\email{fengbo@mail.ihep.ac.cn}
\author{Xinmin Zhang${}^a$}
\email{xmzhang@mail.ihep.ac.cn}
\affiliation{${}^a$Institute of High Energy Physics, Chinese
Academy of Science, P.O. Box 918-4, Beijing 100039, P. R. China}
\affiliation{${}^b$Interdisciplinary Center of Theoretical
Studies, Chinese Academy of Sciences, P.O. Box 2735, Beijing
100080, China} 


\begin{abstract}
Recent released WMAP data show a low value of quadrupole in the
CMB temperature fluctuations, which confirms the early
observations by COBE. In this paper, a scenario, in which a
contracting phase is followed by an inflationary phase, is
constructed. We calculate the perturbation spectrum and show that
this scenario can provide a reasonable explanation for lower CMB
anisotropies on large angular scales.
\end{abstract}

\pacs{98.80.Cq, 98.70.Vc} \maketitle

Recently the high resolution full sky Wilkinson Microwave
Anisotropy Probe (WMAP) data \cite{IEWMAP,Bennett,Verde, Spergel,
Hinshaw} have been released and it is shown that the data is
consistent with the predictions of the standard concordance
${\rm\Lambda}$CDM model. However, there remain two intriguing
discrepancies between WMAP observations and the concordance model.
The data predict a high reionization optical
depth\cite{Kogut}\cite{Fukugita} and a running of the spectral
index\cite{Spergel}, as claimed by WMAP team. The need of a
running has been studied widely\cite{Wang,Lewis,Seljak,fitWMAP}
and many inflation models with large running of a spectral index
have been built \cite{feng,running}. Another surprising
discrepancy comes from the low temperature-temperature(TT)
correlation quadrupole, which has previously been observed by
COBE\cite{cobe}. It is pointed out by Ref.\cite{Lewis} that there
might be some connection between the need for running of the
spectral index and the suppressed CMB quadrupole, and the
significance of the low multipoles has been discussed widely in
the literature\cite{small00}.

Several possibilities to alleviate the low-multipoles problem have
been discussed in the
literature\cite{small01,Linde,small02,fengbo2}. One
straightforward way is to build suppressed primordial spectrum on
the largest scales\cite{Lewis}. This can also lead to other
observable consequences\cite{Kamionkowski,small03}. In the
framework of inflation, changing the inflaton potential and the
initial conditions at the onset of inflation have been proposed
\cite{Linde}. For the latter case, the inflaton has to be assumed
in the kinetic dominated regime initially. Since there are no
primordial perturbations exiting the horizon in such a phase, the
inflation\cite{fengbo2} or contracting phase before kinetic
domination should be required.

In this paper we consider a scenario where a contracting is
followed by an inflationary phase and study its implications in
suppressing CMB quadrupole. For a contracting phase with a kinetic
domination, the primordial perturbations exiting the horizon can
be obtained similar to that of Pre Big Bang (PBB) scenario
\cite{GV}(for a review see \cite{V}). The PBB scenario is regarded
as an alternative to the inflation scenario, but its spectrum is
strongly blue and does not provide the nearly scale-invariant
perturbation spectrum implied by the observations by the evolution
of background field. In the literature there are some proposals of
alternatives for seeding the nearly scale-invariant spectrum in
the contracting phase.  In addition to the ekpyrotic/cyclic
scenario \cite{KOS}, there is a possibility to seed a
scale-invariant spectrum \cite{FB} in which the pressureless
matter is used. For the expanding phase, in addition to the usual
inflation scenario, a slowly expanding phase may also be feasible
\cite{PZ}. In general the cut-off of primordial power spectrum
\cite{Lewis} may indicate a matching between different phases
during the evolution of the early universe.

In this paper we will calculate the perturbation spectrum in the
model with a contracting phase followed by an inflation and fit it
to the WMAP data. Our results show that this scenario can provide
a reasonable explanation for the observed low CMB anisotropies on
large angular scales.

Consider a generic scalar field with lagrangian \be {\cal L}=
-{1\over 2}(\partial_\mu \varphi)^2 -V(\varphi) .\ee For the
spatially homogeneous but time-dependent field $\varphi$, the
energy density $\rho$ and pressure $p$ can be written respectively
as \be \rho ={1\over 2}{\dot \varphi}^2 + V(\varphi) ,~~~
p={1\over 2}{\dot \varphi}^2-V(\varphi). \label{p} \ee The
universe, described by the scale factor $a(t)$, satisfies the
equations \be h^2 = {8\pi G\over 3} \left({1\over 2 }{\dot
\varphi}^2 +V(\varphi)\right) , \label{da} \ee and the
equation of motion of the scalar field is \be {\ddot \varphi} +3h{\dot
\varphi}
+ V^\prime(\varphi)=0 \label{drho} ,\ee where $h ={{\dot a}\over
a}$ is the Hubble parameter.

For the universe in the contracting phase, we have $h<0$. In this case,
$3h{\dot \varphi}$ is anti-frictional, and instead of damping the
motion of $\varphi$ in the expanding phase it accelerates the
motion of $\varphi$. Thus if the time is long enough, a scalar
field initially in a flat part of the bottom of the potential
will roll up along the potential. During this process, \be {1\over
2}{\dot \varphi}^2 \gg V(\varphi), \ee and \be {\ddot
\varphi}+3h{\dot \varphi}\simeq 0 . \ee To match our observational
cosmology, one requires a bounce from the contracting phase to the
expanding phase. In the literature there have been several
proposals for such a nonsingular scenario with the realization of
 the bounce, for instance,
from a negative energy density fluid \cite{F} or the curvature
term \cite{TT} around the transition, or some higher order terms
stemming from quantum corrections in the action \cite{AGN, GMV}.
After the bounce, since $h>0$, $3h{\dot\varphi}$ becomes
frictional and serves as a damping term. Thus the motion of $\varphi$
decays quickly. When the velocity of $\varphi$ is 0, it reverses
and rolls down along the potential driven by $V^\prime(\varphi)$,
and enters the slow-roll regime in which the universe is dominated
by the potential energy of the scalar field \be {1\over 2}{\dot
\varphi}^2 \ll V(\varphi) ,\ee and \be 3h{\dot \varphi} +
V^\prime(\varphi) \simeq 0  .\ee

In general there exist two regimes in this scenario \footnote{A
similar scenario have been proposed \cite{KSS} in which the form
$\sim \varphi^n$ of the potential has been studied numerically and
two regimes, {\it i.e.} $p=\rho $ for the contracting phase and
$p=-\rho$ for the expanding phase, have been found.}. For the
regime before the bounce, the equation of state of the background
is $p\simeq \rho$, consequently we have \be a(t)\sim (-t)^{1\over
3} ,\ee while for the slow-roll regime after the bounce, $p\simeq
-\rho$, so the evolution of the scale factor is given by \be
a(t)\sim \exp{(t)} .\ee For convenience of the calculations on the
perturbation spectrum, we define $dt =a d\eta$ where $\eta$ is the
conformal time. For both phases, we have \be a(\eta)\sim
(-\eta)^{1\over 2} ,~~~~~~  a(\eta)\sim (-\eta)^{-1} \ee and \be
{\cal H} ={a^\prime \over a} ,\ee where the prime denotes the
derivative with respect to $\eta$. For simplify, we neglect the
details of the bounce and focus on an instantaneous transition
between a kinetic-dominated contracting phase and  a nearly de
Sitter phase.
 We set $\eta=0$ and
$a=1$ at the moment of transition for the matching, thus we have
\be a\simeq \sqrt{1-2{\cal H}_0\eta}~, ~~~~~\eta\leq 0 \label{leq}
\ee \be a\simeq {1\over 1-{\cal H}_0\eta}~, ~~~~~\eta\geq 0
\label{geq} \ee where ${\cal H}_0$ is the physical Hubble constant
during the inflationary phase.

Now we study the metric perturbations of the model.
Working in the
longitudinal gauge the scalar perturbations responsible for
 the observed large angle CMB temperature anisotropies
can be written as \cite{MFB} \be ds^2 =
a^2(\eta) (-(1+2\Phi)d\eta^2 +(1-2\Phi) \delta_{ij}dx^i dx^j )
,\ee where $\Phi$ is the Bardeen potential \cite{B}. For the
Mukhanov-Sasaki variable\cite{M}, one has \be v\equiv a\left(\delta
\varphi +{\varphi^\prime\over {\cal H}}\Phi\right)\equiv z\zeta
,\ee where $\varphi$ is the background value of the scalar field
and $\delta \varphi$ denotes the perturbations of the scalar field
during the periods of both phases, contraction and inflation, and $\zeta$
is
the curvature perturbation on uniform comoving hypersurface,
$z\equiv {a\varphi^\prime\over {\cal H}}$. In the momentum
space, the equation  of motion of $v_k$ is \be
v_k^{\prime\prime}+\left(k^2 -{z^{\prime\prime}\over z}\right)v_k
=0  .\ee
 For the contracting phase
before inflation, \be {z^{\prime\prime}\over z}\simeq
{a^{\prime\prime}\over a} \simeq {-{\cal H}_0^2\over (1-2{\cal
H}_0 \eta)^2} . \ee When $k^2\gg {z^{\prime\prime}\over z}$, the
fluctuations are in their Minkowski vacuum, which corresponds to
\be v_k\sim {1\over \sqrt{2k}} e^{-ik\eta} ,\ee thus \be
v_k(\eta)=\sqrt{\pi(1-2{\cal H}_0\eta)\over
8{\cal_H}_0}H_0^{(2)}\left(-k\eta+{k\over 2{\cal H}_0}\right)
 ,\ee where $H_0^{(2)}$ is the second kind of Hankel function
with $0$ order. For the nearly de Sitter phase, \be
{z^{\prime\prime}\over z}\simeq {a^{\prime\prime}\over a} \simeq
{2{\cal H}_0^2 \over (1-{\cal H}_0\eta)^2} ,\ee thus \ba
v_k(\eta)& &=\sqrt{-k\eta +{k\over {\cal H}_0}}\nonumber \\ &
&\left(C_1 H_{3\over 2}^{(1)}(-k\eta +{k\over {\cal H}_0})+C_2
H_{3\over 2}^{(2)}(-k\eta +{k\over {\cal H}_0})\right) \label{vki}
,\ea where $H_{3\over 2}^{(1)}$ and $H_{3\over 2}^{(2)}$ are the
first and second kind of Hankel function with ${3\over 2}$ order
respectively, $C_1$ and $C_2$ are $k$-dependent functions, which
are determined by the matching conditions between two phases.

In general, the details of the dynamics governing the bounce
determines the matching conditions for the calculations of the
spectrum, which specifically depends on whether the curvature
perturbation $\zeta$ on uniform comoving hypersurface or the
Bardeen potential $\Phi$ passes regularly through the bounce
\cite{CDC}( see also \cite{D, TT, BF, PP}). For a bounce scenario
like PBB with higher order correction terms, it has been shown to
the first order in $\alpha^\prime$\cite{CHC, TBF} on the
continuity of the induced metric and the extrinsic curvature
crossing the constant energy density matching surface between the
contracting and the expanding phase, {\it i.e.} $\zeta$ (thus $v$)
passes regularly through the transition. From the matching
condition at the transition point $\eta =0$, {\it i.e.} the
continuity of $v$ and $v^\prime$ implies that
 \ba C_1 =&
&\sqrt{\pi\over 32{\cal H}_0}e^{-ik\over {\cal H}_0} ((1-{2{\cal
H}_0^2\over k^2}-{2{\cal H}_0\over
k}i)H_0^{(2)}\left({k\over 2{\cal H}_0}\right)\nonumber \\
& & +({{\cal H}_0\over k} +i)H_1^{(2)}\left({k\over 2{\cal
H}_0}\right)) ,\label{c1}\ea \ba C_2 &=& \sqrt{\pi\over 32{\cal
H}_0}e^{ik\over h} ((1-{2{\cal H}_0^2\over
k^2}+{2{\cal H}_0\over k}i)H_0^{(2)}\left({k\over 2{\cal H}_0}\right)\nonumber\\
& &+({{\cal H}_0\over k} -i)H_1^{(2)}\left({k\over 2{\cal
H}_0}\right)) ,\label{c2}\ea where $H_0^{(2)}$ and $H_1^{(2)}$ are
the second kind of Hankel function with $0$ and $1$ order
respectively. The spectrum of tensor perturbation is \be P_g =
{k^3\over 2\pi^2}|{v\over a}|^2 \label{p} ,\ee for $\eta
\rightarrow 1/{\cal H}_0$. Substituting (\ref{vki}), (\ref{c1})
and (\ref{c2}) into (\ref{p}), we obtain \be P_g= {{\cal
H}_0^2\over 2\pi^2}k|C_1 -C_2|^2. \label{ps}\ee Since the spectrum
freezes during slow-rolling inflation, the scalar spectrum can be
obtained via the consistency condition $P_s= P_g / r$, where $r$
is a constant. We made a numerical check and find this is a good
approximation.
\begin{figure}[t]
\begin{center}
\includegraphics[width=8cm]{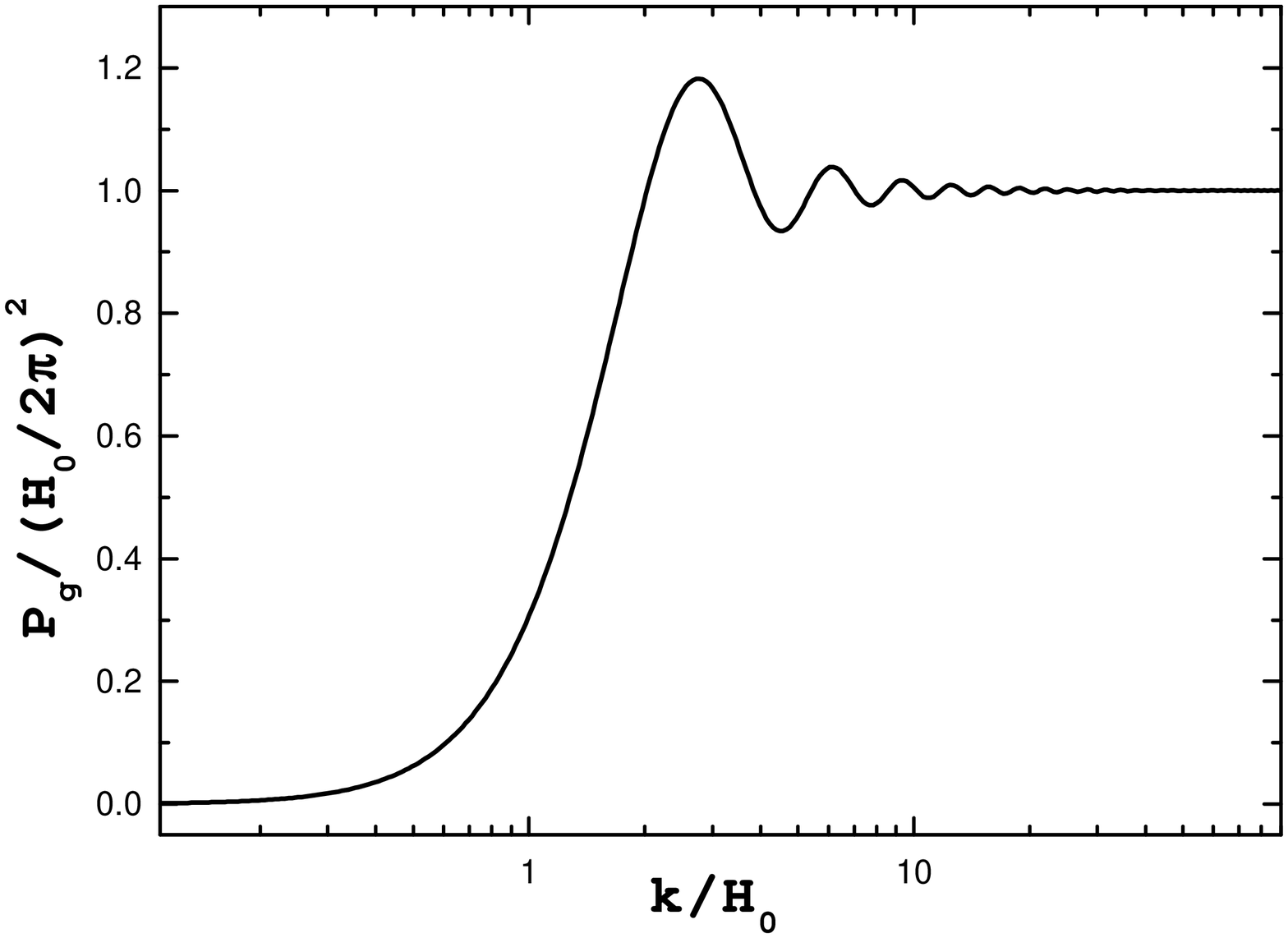}
\caption{The power spectrum ${\cal P}_g$as a function of ${k\over
{\cal H}_0}$. The x-axe is ${k\over {\cal H}_0}$, and the y-axe is
$P_g /\left({{\cal H}_0\over 2\pi}\right)^2$.} \label{scales}
\end{center}
\end{figure}

For $k\ll {\cal H}_0$, the Hankel function can be expanded in term
of a large variable, thus we have approximately \be P_s \sim k^3
\ee on large scale, which is the usual result of PBB scenario.
For $k \gg {\cal H}_0$, the Hankel function can be expanded in
term of a small variable, thus we obtain \be P_s \sim k^0\ee on
small scale, which is the result of inflation scenario.
This is because the large $k$ modes are inside the horizon during
the contracting phase and are not quite insensitive to the
background at this stage. Thus when they cross the horizon during
inflation after the transition, the nearly scale-invariant
spectrum can be generated by the evolution of the background
during inflationary phase. In Fig.1 we plot $P_g$ in (\ref{ps})
as a function of ${k\over {\cal
H}_0}$. We see that for $k\sim {\cal H}_0$ the amplitude of the
spectrum oscillates and for $k\ll {\cal H}_0$ it decreases rapidly
and gets a cutoff. Therefore for an appropriate choice of the
e-folds number of inflation, it is possible to suppress the lower
multipoles of the CMB anisotropies.

\begin{figure}
\includegraphics[scale=0.3]{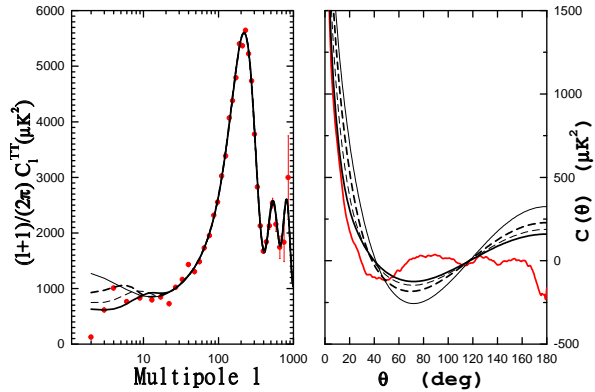}
\caption{\label{Cl} CMB anisotropy and two-point temperature
correlation function for  the scale invariant spectrum and the
spectrum with a cutoff. Left: From left top to bottom, the lines
stand for scale invariant spectrum, spectrum with a cutoff with
${\cal H}_0=2.1,3.1$ and $4.1\times 10^{-4}$ Mpc$^{-1}$. Other
parameters are fixed at $h=0.73$, $\Omega_b h^2=0.023$,
$\Omega_{cdm} h^2= 0.117 $ and $\tau=0.2$. Right: From right top
to bottom,the lines stand for scale invariant spectrum, spectrum
with a cutoff with ${\cal H}_0=2.1,3.1$ and $4.1\times 10^{-4}$
Mpc$^{-1}$ and the WMAP released data.}
\end{figure}

Now we fit the resulting primordial spectra to the current WMAP TT
and TE data. In our model the sufficient contraction makes the
universe flat, so we take $\Omega_k =0$. We vary grid points with
ranges $[0.65,0.75]$, $[0.021,0.024]$, $[0.12,0.16]$, $[0.05,0.3]$
and $[0,0.001]$ Mpc$^{-1} $ for $h$, $\Omega_b h^2$, $\Omega_{cdm}
h^2$, $\tau$ and ${\cal H}_0$ respectively.  At each point in the
grid we use subroutines derived from those made available by the
WMAP team to evaluate the likelihood with respect to the WMAP TT
and TE data \cite{Verde}. The overall amplitude of the primordial
perturbations has been used as a continuous parameter. Tensor
contribution has not been considered since $r$ can be very small.
We get a minimum $\chi^2=1428.2$ at $h=0.73$, $\Omega_b
h^2=0.024$, $\Omega_{cdm} h^2= 0.116 $, $\tau=0.2$ and ${\cal
H}_0=2.0\times 10^{-4}$ Mpc$^{-1}$. We also run a similar code for
the scale invariant spectrum for comparison and get a minimum
$\chi^2=1429.7$ at $h=0.73$, $\Omega_b h^2=0.024$, $\Omega_{cdm}
h^2= 0.116 $ and $\tau=0.2$. This means our primordial spectrum is
favored at $>1.1\sigma$ than the scale invariant spectrum in our
realization. ${\cal H}_0$ can be given in our fit with ${\cal
H}_0\lesssim 5.0\times 10^{-4}$ Mpc$^{-1}$. However as we have set
$a=1$ at the transition scale instead of today, the exact physical
 energy scale during the transition cannot be known due to the
 uncertainty in the number of e-folding and details
 of reheating\cite{Lyth reports,fgw}.
  In Fig. \ref{Cl} we show the resulting
CMB TT multipoles and two-point temperature correlation function
for the scale invariant spectrum and our spectrum with a cutoff in
our parameter space. One can see that the resulting CMB TT
quadrupole and the correlation function at $\theta\gtrsim 60^\circ
$ can be much better suppressed for spectrum with a cutoff  than
in the scale invariant case. It is noteworthy that the uncertainty
by cosmic variance plays an significant role around the smallest
CMB multipoles, which is much larger than WMAP's instrumental
noise. WMAP team predicts an extremely low TT quadrupole $\delta
T_2 =123 \mu K^2$. Meanwhile the best fit power law and
running-spectral index $\Lambda$CDM model predict $\delta T_2
=1107 $ and 870 $\mu K^2$ respectively\cite{IEWMAP}. Our cutoff
spectrum can give $\delta T_2$ as low as 620 $\mu K^2$. It is not
yet compatible with WMAP quadrupole within cosmic variance limit
since the lowest $\delta T_2$ is $620\times(1-\sqrt{2/5})\sim 228$
$\mu K^2$. However as claimed by Efstathiou\cite{GE0310} the
pseudo-$C_l$ estimator used by the WMAP team might be non-optimal
and the quadrupole  is found to lie between 176 and 250 $\mu K^2$
and more likely to be at the upper bound of the range. Thus our
model can be actually workable and future WMAP data may present a
more presice check.

In summary, we construct a scenario in which a contracting phase
is matched to an inflationary phase instantaneously. We calculate
the spectrum of the scalar perturbation and find that the power
spectrum on large scale is suppressed due to $\sim k^3$, which is
the usual result of PBB scenario, and on small scale the nearly
scale-invariant spectrum of inflation is recovered. Thus our
scenario can provide a reasonable explanation for lower CMB
anisotropies on large angular scales. Although in our proposed
scenario, we neglect the physical details of the bounce, the
results obtained by us reflect the generic feature of model in
which the inflation phase follows the contracting phase of PBB. In
our scenario, we not only obtain the suppressed lower multipoles,
which is connected with the physical detail of PBB and bounce, but
also avoids the initial singularity by the bounce. Furthermore,
our scenario makes an attempt to improve the PBB scenario on the
graceful exit problem with a period of inflation, which is worth
studying further.

\begin{acknowledgments}
We thank Robert Brandenberger, Qing-Guo Huang and Mingzhe Li for
helpful discussions. We acknowledge the using of CMBFAST
program\cite{cmbfast,IEcmbfast}. This work is supported by the
National Natural Science Foundation of China under the grant No.
10105004, 19925523, 10047004 and also by the Ministry of Science
and Technology of China under grant No. NKBRSF G19990754.
\end{acknowledgments}

\newcommand\PRPT[3]{~Phys.Rept.{\bf ~#1}, #2~(#3)}

\end{document}